\pdfoutput=1
\documentclass[aps,twocolumn,superscriptaddress]{revtex4-1}

\usepackage{graphicx} % Allows for eps images
\usepackage{epsfig}
\usepackage{amsmath} % AMS Math Package
\usepackage{amsthm} % Theorem Formatting
\usepackage{amssymb}	% Math symbols such as \mathbb
\usepackage{physics}
\usepackage{graphics} % or graphicx 
\usepackage{hyperref} %clickable references.
\hypersetup{
    colorlinks,
    citecolor=blue,
    filecolor=black,
    linkcolor=red,
    urlcolor=blue
}
\usepackage[normalem]{ulem}
\usepackage[english]{babel}
\usepackage{blindtext}

\begin{document}

\title{The hidden topological structure of flow network functionality}
\author{Jason W. Rocks}
\author{Andrea J. Liu}
\author{Eleni Katifori}
\affiliation{Department of Physics and Astronomy, University of Pennsylvania, Philadelphia, PA 19104, USA}

\begin{abstract}

The ability to reroute and control flow is vital to the function of venation networks across a wide range of organisms. 
By modifying individual edges in these networks, either by adjusting edge conductances or creating and destroying edges, 
organisms can robustly control the propagation of inputs to perform specific tasks.
However, a fundamental disconnect exists between the structure and function of these networks: 
networks with different local architectures can perform the same functions. 
Here we answer the question of how structural changes at the microscopic level are able to collectively create functionality at the scale of an entire network.
Using persistent homology, we analyze networks tuned to perform complex multifunctional tasks.
We find that the responses of such networks encode a hidden topological structure composed of sectors of uniform pressure.
Although these sectors are not apparent in the underlying network architectures, 
we find that they nonetheless correlate strongly with the tuned function.
We conclude that the connectivity of these sectors, rather than that of the individual nodes, 
provides a quantitative relationship between structure and function in flow networks.
Finally, we use this topological description to place a bound on the limits of task complexity.

\end{abstract}
\maketitle

Many biological fluid transport networks can redirect flow as dictated by the needs of the system.
For example, by dynamically contracting and dilating blood vessels, the cerebral vasculature actively controls blood flow to support local neuronal activity~\cite{Cipolla2016, Gao2015};
impairment of this ability has been linked to neurological diseases~\cite{Sweeney2018} such as Alzheimer's disease~\cite{Liesz2019}.
More generally, the ability to tune the conductances of edges or locally restructure connectivity enables animals~\cite{Tuma2008, Meigel2019},  
plants~\cite{Pittermann2010, Sack2013}, fungi~\cite{Heaton2012}, and slime molds~\cite{Tero2008} 
to control the spatial distribution of water, nutrients, oxygen, or metabolic byproducts.

Recently, Rocks, Ronellenfitsch, et al.~\cite{Rocks2019} demonstrated that flow networks are remarkably tunable with the ability to perform highly complex multifunctional tasks.
They showed that the pressure differences at a large number of pre-specified sites in a flow network can be simultaneously controlled by selectively tuning edge conductances.
In other words, by modifying the local (or microscopic) structure of a network, 
it is possible to attain a pre-specified collective property or ``function," 
namely, a desired pressure drop at a collection of specific target edges.  
Moreover, different networks can easily be tuned to develop the same function (i.e. the same number of target edges and desired target pressure differences).
These results raise a fundamental question: How do changes in the local structure of a flow network collectively achieve specific functions? 
More simply, what is the relationship between structure and function?

Here we use persistent homology to identify the underlying basis of function in flow networks. 
We find that the structure-function relationship is \emph{topologically} encoded in the response:
as a network is tuned to achieve a desired target pressure difference at a number of different sites, 
it separates into distinct sectors of relatively uniform node pressure,
even as the underlying network architecture remains a single connected component.
It is the connectivity, or topology, of these  \textit{sectors} that determines the function, rather than that of the actual nodes.
Our finding provides a simple, unifying topological description of all networks tuned for the same function,
regardless of the underlying network architecture, along with the quantitative means to compare networks tuned for different functions.
We demonstrate that this description is robust even when the magnitude of the tuned response is small and the sectors cannot be identified by eye.
We also use this description to provide structural insight into the limits of multifunctionality (maximum number of tunable target sites).

To identify the structure-function relationship, we create ensembles of flow networks that each perform the same function. 
To accomplish this, we first generate a collection of networks and then tune each one by adjusting the conductances of its edges~\cite{Rocks2019}. 
More specifically, we consider flow networks (or equivalently, resistor networks) in which edges between nodes represent pipes (linear resistors).
In this framework, the response of a network to external stimuli, described by a set of pressures (voltages) on the nodes,
is governed by a discrete version of Laplace's equation that is equivalent to Kirchoff's laws.
We use contact networks of randomly-generated two and three-dimensional packings 
of soft spheres with periodic boundary conditions, created using standard jamming algorithms. 
To derive a flow network from a contact network, we assign to each edge a conductance value, chosen randomly from the range $0.1$ to $1.0$ in discrete increments of $0.1$.

Next, we tune each flow network to perform a specific function, so that the pressure differences of a specified set of target edges respond by at least an amount $\Delta$
(chosen to be non-negative) when a unit pressure difference is applied across a specified ``source" edge.
For each network in the ensemble, the source and target edges are chosen at random with the constraint that they do not share any nodes.
To achieve a target pressure difference $\Delta p_T \ge \Delta$ across each target edge,
we use a greedy algorithm: in each step we increase or decrease the conductance of a single edge by $0.1$ 
(staying within the range $0$ to $1$, inclusively), modifying the edge conductance that best optimizes the total response at that step 
(for further details concerning network generation and tuning, see with Refs.~\cite{Rocks2019} and ~\cite{Rocks2019b}, 
along with similar earlier work on mechanical networks in Ref.~\cite{Rocks2017}). 
We note that the details of the tuning algorithm do not affect the generality of the structure-function relationship we identify.

\begin{figure}[t!]
\centering
\includegraphics[width=1.0\linewidth]{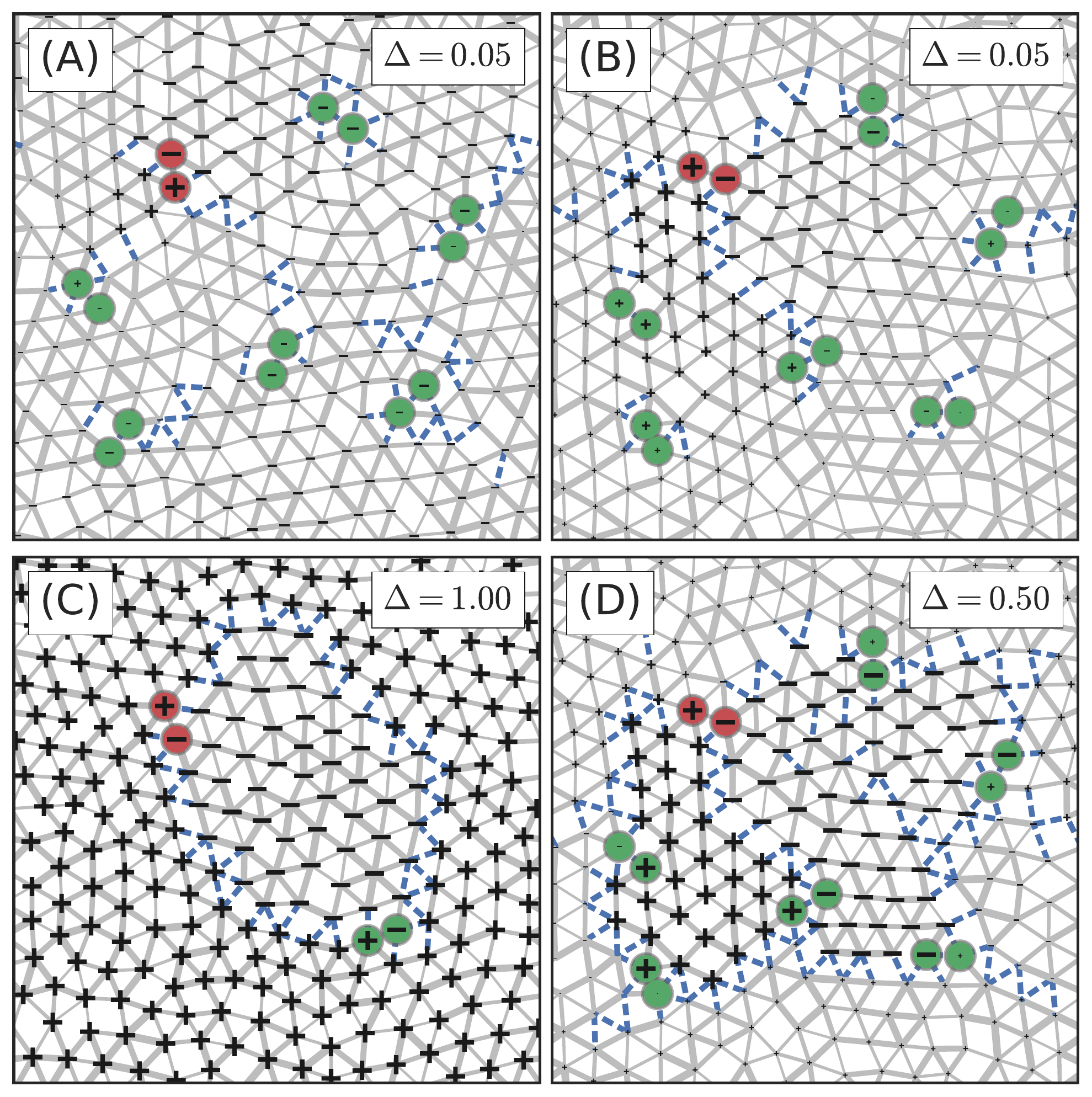} 
\caption{(A), (B) Comparison of a pair of two-dimensional flow networks that perform the same function whose structures differ both before and after tuning.
In both examples, when a unit pressure difference is applied across the source nodes (shown in red), six targets, each composed of a pair of nodes (shown in green), respond with a pressure difference of at least $\Delta = 0.05$. 
For visual clarity, similar positions have been chosen for the sources and targets.
Node pressures are in black, with symbol type denoting sign and symbol size denoting magnitude. 
Edge thickness corresponds to the conductance with thick dashed blue lines indicating edges that have been entirely removed (set to zero conductance) in the tuning process.
(C) A network with a single function tuned to a maximum pressure difference of $\Delta = 1.0$. 
(D) The network in (B) tuned to $\Delta = 0.5$ instead of $\Delta=0.05$.}
\label{fig:struct_comp}
\end{figure}

Figs.~\ref{fig:struct_comp}(A) and (B) illustrate the discrepancy between structure and function: two different networks are tuned to perform the same task, 
namely to have six target edges each tuned to the same target pressure difference of $\Delta = 0.05$ relative to the source (we have also chosen similar relative positions of the source and targets for visual clarity).
Clearly, the spatial distributions of edge conductances (indicated by edge thickness) and pressures (indicated by the size of the symbols showing the sign of the pressure on each node) in the networks are noticeably different; 
it is unclear from Figs.~\ref{fig:struct_comp}(A) and (B) whether the underlying structures of the two tuned networks share anything in common.

We gain insight by examining the networks when the targets are tuned to larger pressure differences.
Fig.~\ref{fig:struct_comp}(C) displays a network with a single target tuned to the extreme limit $\Delta = 1$, 
the maximum achievable pressure difference at a target edge.
Here the network clearly separates into two distinct sectors of perfectly uniform node pressure, 
connected only by a single edge between the source nodes.
These two sectors are separated by a crack-like structure with pressure differences of precisely $1.0$ 
across edges that have been removed during the tuning process, denoted by dashed blue edges. 
Similarly, Fig.~\ref{fig:struct_comp}(D) displays the network from Fig.~\ref{fig:struct_comp}(B), but with each target edge tuned to $\Delta = 0.50$. 
In this extreme case, the network has separated into three distinct sectors, each of almost perfectly uniform pressure.
In Figs.~\ref{fig:struct_comp}(C) and (D), the creation of the desired function is purely due to the creation of these sectors.
Almost all edges (except for those at the source) connecting the different sectors are removed, so that each sector comprises a different connected component.
The exact details of the local structure (which specific edges are modified) do not matter as long as this partitioning takes place.
In these extreme cases, the relationship between structure and function is clear:
the increase in the number of connected components is directly tied to the function. 
The emergence of these sectors represents a topological change in the overall network connectivity beyond that of the local edge structure.
Clearly, this description extends to all networks tuned to this extreme limit.

For smaller $\Delta$, as in Figs.~\ref{fig:struct_comp}(A) and (B), the entire network remains highly interconnected--there is only one connected component even after the desired function is achieved. 
The challenge is therefore to apply the insight gained from the extreme case to smaller $\Delta$.

\begin{figure*}[t!]
\centering
\includegraphics[width=1.0\linewidth]{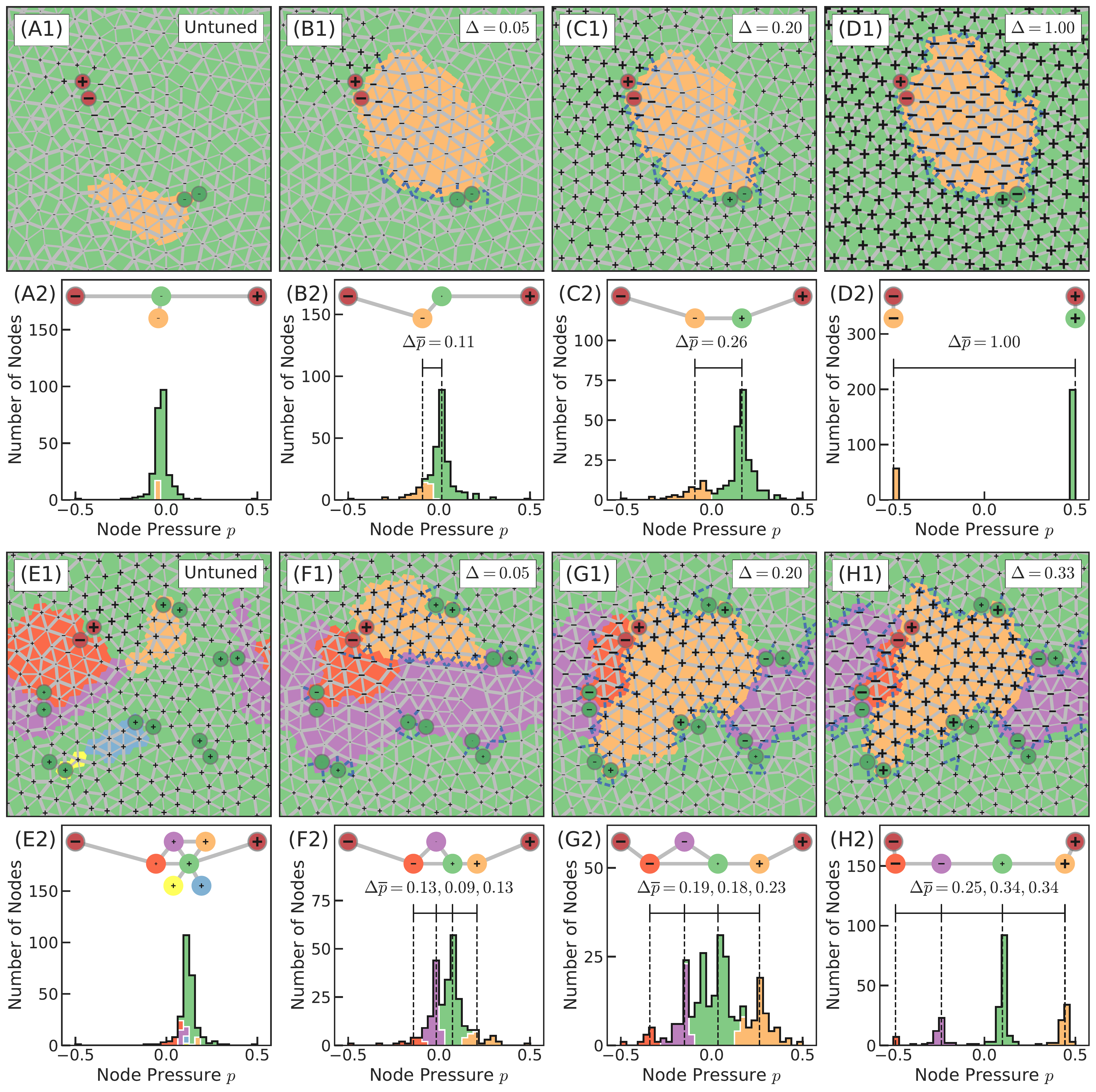}
\caption{Topological structure of the response for the network of Fig~\ref{fig:struct_comp}(C) with a single target (A) before tuning and tuned for target pressure differences of (B) $\Delta = 0.05$ (C) $\Delta = 0.2$, and (D) $\Delta = 1.0$. 
Similarly, a multifunctional network with six separate targets (E) before tuning and tuned for (F) $\Delta = 0.05$ (G) $\Delta = 0.2$, and (H) $\Delta = 0.33$. 
(First and Third Rows) Sectors characterizing the response are highlighted by color. Source nodes are red and target nodes green. 
Black symbols indicate node pressures, where the symbol type denotes sign and size denotes magnitude. 
Edge thickness corresponds to conductance, with thick dashed blue lines indicating fully removed edges. 
(Second and Fourth Rows) Histograms of node pressures colored to indicate contributions from nodes in each of the corresponding sectors in the networks above.
The median node pressure $\overline{p}$ of each sector is shown as a vertical dashed line and
differences in these median pressures $\Delta\overline{p}$ corresponding to neighboring peaks in the histograms are indicated. 
Inset in each histogram is a schematic depicting the connectivity between sectors, 
represented as nodes with source nodes in red. Edges indicate existence of edges between sectors in tuned network.
Symbols (and approximate horizontal position) denote sign and magnitude of median node pressures.
}
\label{fig:sectors}
\end{figure*}

To proceed, we utilize the observation that in the extreme high $\Delta$ case, 
each connected component is comprised of nodes of equal pressures. 
The pressure differences on all edges contained within each sector is zero, while the pressure differences between sectors is nonzero.
The change in the number of connected components is directly linked to the pressure differences at the target edges.
We therefore seek to identify analogous sectors for networks tuned to smaller $\Delta$.
The extreme case suggests that methods derived from topological data analysis might be helpful. 
We use persistent homology, an analysis that discerns topological features in topologically and/or geometrically structured data~\cite{Edelsbrunner2010, Otter2017}. 
A benefit of this technique is that it provides a systematic means of identifying topological features at all scales encoded in a function (the pressure response) defined in some space (the network). 
Each feature we identify corresponds to a region of relatively small pressure differences (relatively uniform node pressures) on the edges in the network. 
However, many of these regions owe their existence to small spatial fluctuations in the pressure response, and their importance to the function is unclear.
This is where a second benefit of persistent homology comes into play: each identified feature is assigned a measure of significance, called \textit{persistence}.
We use these persistence values to perform topological coarse-graining (a form of hierarchical clustering), 
combining as many of the lowest persistence regions with their neighbors as possible to achieve the smallest number of sectors.
However, since the functions we tune into the networks require creating pressure differences between each pair of target nodes,
we avoid combining regions that would place both nodes that comprise a single target edge into the same sector (details in Ref.~\cite{Rocks2019b}).
This technique results in a set of sectors that are minimal, as significant as possible, and correlated to the tuned response.

\begin{figure}[t!]
\centering
\includegraphics[width=1.0\linewidth]{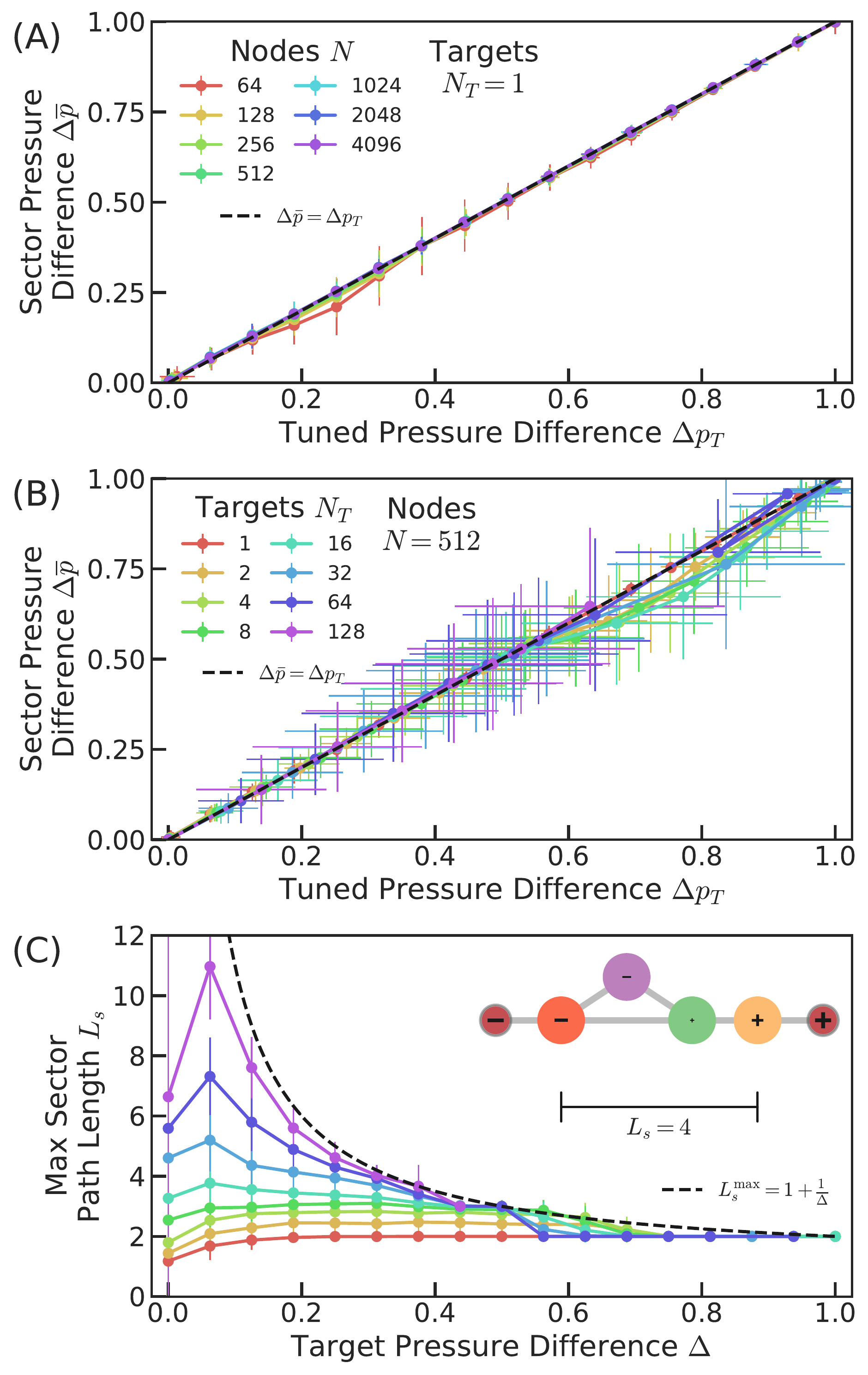}
\caption{Correlation of pressure difference between sectors $\Delta\overline{p}$ and tuned pressure difference of each target $\Delta p_T$ for three-dimensional networks of various sizes $N$ (number of nodes) and numbers of target edges $N_T$.
(A) Correlation as a function of system size for $N_T=1$. 
and (B) as a function of the number of targets for $N=512$.
Both the $\Delta p_T$ and $\Delta\overline{p}$ values for each point are averaged over up to 256 independent networks tuned to the same value of $\Delta$. 
Error bars represent standard deviations.
The diagonal black dashed lines indicate a perfect correlation ($\Delta\overline{p} = \Delta p_T$).
(C) Average maximum path length $L_s$ of monotonically decreasing sector pressure between the two source nodes, measured in terms of the number of sectors.
This path length closely abides by the upper bound set by $L_s^{\max}$. Plot colors indicate the same information as legend in (B). 
(C-Inset) Schematic of connectivity between sectors for network in Fig.~\ref{fig:sectors}(G1) with $\Delta = 0.2$.
Nodes correspond to sectors with source nodes in red and edges indicate existence of edges between sectors in tuned network.
Symbols denote sign and magnitude of median node pressures, with nodes positioned from left to right in order of increasing pressure.
The maximum path length is $L_s=4$, less than the maximum value of $L_s^{\max} = 6$ for $\Delta = 0.2$.
}
\label{fig:seg_pres_stats}
\end{figure}

Figs.~\ref{fig:sectors}(A-D) show how the resulting sectors evolve with $\Delta$ for the case of a single target,
while Figs.~\ref{fig:sectors}(E-H) show the evolution of the sectors for a multifunctional network.
The networks segregate into multiple sectors (two and four, respectively), each composed of nodes with relatively uniform pressures.
Fig.~\ref{fig:sectors} also depicts histograms of the node pressures for each network, colored according to their corresponding sectors.
Depicted above each histogram is schematic of the connectivity between sectors, representing the coarse-grained topology of each network.

The identified sectors quantitatively characterize the tuned function.
To show this, we measure the median node pressure $\overline{p}$ of each sector, shown as vertical dashed lines in the histograms in Fig.~\ref{fig:sectors}.
For any pair of sectors, we can measure the difference in these median node pressures, which we call the sector pressure difference $\Delta \overline{p}$.
We observe that the value of $\Delta \overline{p}$ measured between a pair of sectors corresponding to neighboring peaks in a histogram often corresponds closely to the desired target pressure difference $\Delta$.
Therefore, for each pair of target nodes, we measure $\Delta \overline{p}$ between their corresponding sectors
and compare this value to the actual pressure difference between the target nodes, $\Delta p_T$, where $\Delta p_T$ is tuned to satisfy $\Delta p_T \ge \Delta$.
Fig.~\ref{fig:seg_pres_stats} shows the correlation between $\Delta \overline{p}$ and $\Delta p_T$ for each target for various system sizes $N$ and numbers of targets $N_T$.
For this analysis, we present results for three-dimensional networks (results for two-dimensional networks are presented in Ref.~\cite{Rocks2019b}).
We see that on average, $\Delta \overline{p}$ is almost perfectly correlated with $\Delta p_T$ for every system size and number of targets. 
We see that for larger networks and smaller numbers of targets the spread of the distributions around each point (standard deviation indicated by error bars) is extremely small.
Figs.~\ref{fig:seg_pres_stats}(A) and (B) show that the identified sectors robustly capture the network structures corresponding to the tuned response.

The limit of multifunctionality in flow networks is governed by a constraint-satisfaction phase transition~\cite{Rocks2019}. 
As the number of targets increases, there is a transition from a regime in which the response of each target can always be satisfied to one where not all responses can be satisfied.
Here we establish an approximate upper bound on the number of sectors in a network tuned to a pressure difference of at least $\Delta$ at multiple targets. 
Consider a sequence of sectors connected in series from the higher-pressure source node to the lower-pressure source node in each network, such that $\overline{p}$ decrease monotonically.
By Kirchoff's voltage rule, the sum of pressure differences along such a path cannot exceed the source pressure difference $\Delta p_S$. 
Assuming that at least one pair of target nodes straddles each pair of neighboring sectors so that their sector pressure difference are at least $\Delta\overline{p} \geq \Delta$,
the maximum number of sectors, or maximum path length, in the longest such sequence is
\begin{align}
L_s^{\max} = 1 + \frac{\Delta p_S}{\Delta}.\label{eq:limit}
\end{align}
Fig.~\ref{fig:seg_pres_stats}(C) shows the average observed maximum path length $L_s$ as a function of $\Delta$ for varying numbers of targets $N_T$. 
The inset of Fig.~\ref{fig:seg_pres_stats}(C) shows how the maximum path length is calculated using the schematic of the sector connectivity for the network in Fig.~\ref{fig:sectors}(G1).
As $\Delta$ increases at fixed $N_T$, the number of sectors decreases so that it never almost exceeds $L_s^{\max}$.
We conclude that the maximum number of targets that can be tuned successfully is indirectly constrained by $L_s^{\max}$.  
For certain combinations of target edges, solutions with the required number of sectors in series cannot be found, and the response of every target edge cannot be satisfied.

In summary, we have established a quantitative characterization of function in flow networks by analyzing the structure of their responses using persistent homology.
When a network is tuned to have desired minimum pressure difference at a collection of targets, 
it partitions into sectors of relatively uniform node pressures. 
The difference in the median pressure between a pair of sectors correlates strongly with the response of target edges spanning the sectors.
For small tuned pressure differences, these sectors are not apparent to the eye. 
Nevertheless, they are topologically encoded in the response of the network and constitute the structure that is relevant to the function.

This sector-based picture provides a unifying description for all flow networks tuned to perform this class of functions. 
Although the local node connectivity and geometrical structure can differ between two networks tuned for the same function, 
the commonality in structure of the networks encapsulated by the sector connectivity becomes apparent when viewed through a topological lens.
This leads us to propose a refinement of the structure-function paradigm in the context of functional flow networks. 
Since the tuning process is inherently topological, 
the aspect of structure that relates to function is also topological; 
it is the relationship between the \textit{topological structure of the response} and function that is important. 

The techniques we have demonstrated, along with the resulting characterization of the tuning process, provide the foundations for an new way to analyze and characterize vascular networks. 
Obtaining an accurate and complete map of every single vessel of an entire organ or organism poses a difficult experimental challenge, 
as vasculature networks frequently consist of millions of nodes and span a range of length scales~\cite{DiGiovanna2018}. 
Our work shows that for certain applications, obtaining this complete map of the network connectivity may not be necessary. 
Rather, we suggest that sampling the local \emph{node pressures} on coarser scales may provide more robust information for characterizing the function of vascular systems.

\begin{acknowledgments}
We thank R. D. Kamien and S. R. Nagel for instructive discussions. This research was supported by the NSF through DMR-1506625 (J.W.R.) and PHY-1554887
(E.K.), the Simons Foundation through 454945 (J.W.R. and A.J.L.), 327939 (A.J.L.), and 568888 (E.K.),
and the Burroughs Welcome Career Award (E.K.).
\end{acknowledgments}

\bibliographystyle{unsrt}

%\bibliography{flow_multifunc}

\end{document}